\documentclass[aps,prb,twocolumn,superscriptaddress,amsmath,amssymb]{revtex4-2}

\usepackage{graphicx}
\usepackage{xcolor}
\graphicspath{{figures/}}
\usepackage{siunitx}
\usepackage[normalem]{ulem}

\newcommand{\void}[1]{}

\begin{document}

\title{Floquet State Depletion in AC Driven Circuit QED}

\author{Ming-Bo Chen}
\thanks{These authors contributed equally to this work.}
\author{Bao-Chuan Wang}
\thanks{These authors contributed equally to this work.}
\affiliation{Key Lab of Quantum Information, CAS, University of Science and Technology of China, Hefei, China}
\affiliation{CAS Center for Excellence in Quantum Information and Quantum Physics, University of Science and Technology of China, Hefei, Anhui 230026, China}

\author{Sigmund Kohler}
\email{sigmund.kohler@icmm.csic.es}
\affiliation{Instituto de Ciencia de Materiales de Madrid, CSIC, E-28049 Madrid, Spain}

\author{Yuan Kang}
\author{Ting Lin}
\author{Si-Si Gu}
\author{Hai-Ou Li}
\author{Guang-Can Guo}
\affiliation{Key Lab of Quantum Information, CAS, University of Science and Technology of China, Hefei, China}
\affiliation{CAS Center for Excellence in Quantum Information and Quantum Physics, University of Science and Technology of China, Hefei, Anhui 230026, China}

\author{Xuedong Hu}
\affiliation{Department of Physics, University at Buffalo, SUNY, Buffalo, New York 14260-1500, USA}

\author{Hong-Wen Jiang}
\affiliation{Department of Physics and Astronomy, University of California at Los Angeles, California 90095, USA}

\author{Gang Cao}
 \email{gcao@ustc.edu.cn}
\affiliation{Key Lab of Quantum Information, CAS, University of Science and Technology of China, Hefei, China}
\affiliation{CAS Center for Excellence in Quantum Information and Quantum Physics, University of Science and Technology of China, Hefei, Anhui 230026, China}

\author{Guo-Ping Guo}
\email{gpguo@ustc.edu.cn}
\affiliation{Key Lab of Quantum Information, CAS, University of Science and Technology of China, Hefei, China}
\affiliation{CAS Center for Excellence in Quantum Information and Quantum Physics, University of Science and Technology of China, Hefei, Anhui 230026, China}
\affiliation{Origin Quantum Computing Company Limited, Hefei, Anhui 230026, China}

\begin{abstract}
We perform Floquet spectroscopy in a GaAs double quantum dot system coupled
to a high-impedance superconducting resonator.  
By applying microwave induced consecutive passages under a double resonance condition, we observe novel Landau-Zener-St\"uckelberg-Majorana interference patterns that stem from a cavity-assisted interference pattern modified by the depletion of the ground state.  
Our experimental results reveal the stationary state behavior of a strongly driven two-level system, and are consistent with the simulations based on our theoretical model.  
This study provides an excellent platform for investigating the dynamics of Floquet states in the presence of strong driving.
\end{abstract}

\date{\today}
\maketitle

Driving a quantum two-level system strongly and periodically leads to
a variety of quantum effects, such as
Landau-Zener-St\"uckelberg-Majorana (LZSM) interference and multi-photon
resonance \cite{Shevchenko2010a, Silveri2017a}.  These phenomena provide clear
evidence for the quantum coherence of a physical system
\cite{Oliver2005a, Sillanpaa2006a,Stehlik2012a,Cao2013a, Forster2014a,
Silveri2015a, Gonzalez-Zalba2016a,Bogan2018a,
Mi2018b,Han2019a,Wen2020a}
and can be employed to probe the energy spectrum \cite{Berns2008a,
Ferron2016a} and the Floquet spectrum \cite{Kohler2018a}.
Since driving and active gating are integral parts of the quantum circuit
paradigm for quantum computing, 
a thorough understanding of the strongly driven dynamics of a qubit could help to explore the limits to qubit control and manipulation.

Theoretical investigations of driven dissipative systems yielded interesting
results in recent years.  For example, it has been shown that at low
temperatures, the stationary state of a dissipative driven quantum system
is often dominated by one particular Floquet state.  Depending on the
system parameters and the driving frequency, this may be the state with a
certain quasienergy \cite{Hartmann2017a, Engelhardt2019a} or the one with
the smallest mean energy \cite{Kohler1998a}.  At avoided \cite{Kohler1998a}
or exact \cite{Engelhardt2019a} level crossings of the Floquet spectrum,
however, the stationary state may turn from an (almost) pure state
to a mixed state made up of several Floquet states.  Despite these
theoretical advances in relating Floquet state dynamics to spectral properties
of strongly driven quantum systems, corresponding experimental
investigations \cite{Magazzu2018a} in semiconductor systems are scarce.

Cavity quantum electrodynamics \cite{HarocheRaimond} allows the study of
atom-photon interaction at the single quantum level \cite{Thompson1992a,
Brune1996a, Schuster2010a,Kubo2010a, Teufel2011a,Tabuchi2014a}.  Recent
progress in hybrid circuit quantum electrodynamics (cQED) systems, composed
of high quality microwave resonators and semiconductor double quantum dots
(DQDs), enables similar explorations in these artificial atoms at the
microwave regime.  For example, it has been shown that strong
electron-photon coupling can be realized in such a cQED system
\cite{Stockklauser2017a, Mi2017b, Bruhat2018a,Samkharadze2018a,
Mi2018a,Landig2018a}, and photon-mediated interaction can provide
long-distance coupling between qubits
\cite{Delbecq2013a,vanWoerkom2018a,Borjans2020a,Wang2021a}.  An additional
benefit of a high quality resonator is that it can also act as
a high-sensitivity dispersive photonic probe of the DQD \cite{Petersson2012a, Frey2012a}, as is widely done for superconducting qubits \cite{Blais2004a, Arute2019a}.

In this work, we study the strongly driven dynamics of a two-level system
by performing Floquet spectroscopy in a DQD-resonator hybrid system.  When
we apply a continuous microwave tone onto the DQD, the nonequilibrium
population of the eigenstates displays novel features in the
cavity-assisted LZSM interference patterns measured by the high-impedance superconducting
resonator.  In particular, when the hybrid system is driven near
half of the resonator frequency, there arise in the interference fringes
crescent-shaped holes, where reflectance peaks acquire troughs in analogy to
optical hole-burning \cite{WallsMilburn1995a}.  Applying the theory for dispersive
readout of driven systems \cite{Kohler2018a}, we reveal that this phenomenon
is caused by a redistribution of the Floquet state population at avoided
quasienergy crossings where the system experiences a double resonance, and can be
controlled via the driving parameters.

\begin{figure}[thp]
\includegraphics[width=\columnwidth]{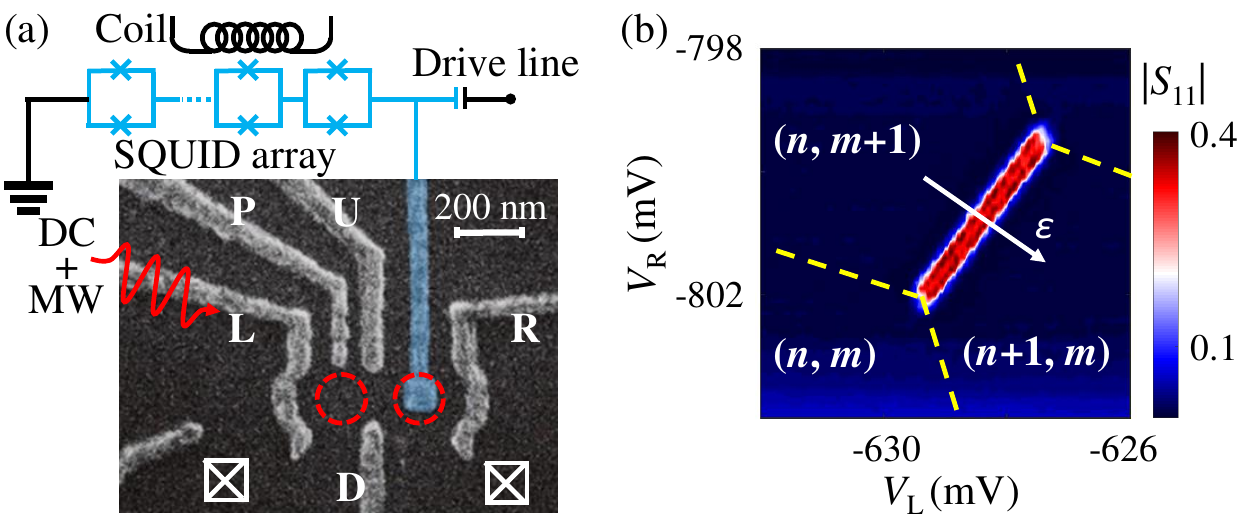}
\caption{(a) Scanning electron micrograph of the device. Two quantum dots
indicated by dotted circles are defined by metallic top gates. The right
plunger gate is connected to the SQUID array resonator.
The qubit is driven by continuous microwave tone via the gate L. (b) Charge
stability diagram of the DQD measured by the resonator.
}
\label{fig:setup}
\end{figure}

Our device consists of a semiconductor DQD and a 1/4
wavelength superconducting quantum interference device (SQUID) array resonator
with a high impedance of $Z_\text{r}\approx\SI{1}{\kilo\ohm}$, as illustrated
in Fig.~\ref{fig:setup}(a). A few electrons are trapped in the DQD defined
electrically within a GaAs/AlGaAs
heterostructure.  The electron occupation $(m,n)$ of the DQD is controlled
by gate voltages $V_\text{L}$ and $V_\text{R}$,
as depicted in the charge stability diagram in Fig.~\ref{fig:setup}(b).  In
the parameter range considered, the DQD dynamics is restricted to the
charge states with one excess electron in the left or right QD, denoted by
$|L\rangle$ and $|R\rangle$, respectively.  Within this basis, the DQD
Hamiltonian can be written as $H_0=\varepsilon \sigma _z/2+t_\text{c}\sigma_x$, where $\varepsilon$ is the detuning between $|L\rangle$ and
$|R\rangle$ controlled by $V_\text{L}$, $t_{\rm c}$ is the interdot
tunnel coupling, while $\sigma_z$ and $\sigma_x$ are Pauli matrices.
Diagonalizing $H_0$, we obtain the energy splitting of a free qubit, $\Delta E=\sqrt{\varepsilon^2+(2t_c)^2}$.

The DQD is connected via the right plunger gate [blue in Fig.~\ref{fig:setup}(a)] to a high-impedance SQUID array resonator with a tunable resonance frequency $\nu_\text{r}$.
Unless specified otherwise, $\nu_\text{r}=\SI{6.51}{\GHz}$.  The internal loss
rate, external coupling rate and total linewidth of the resonator are
$(\kappa _\text{i},\kappa_\text{e},\kappa) /2\pi
=(31.0,34.2,65.2)\mathrm{MHz}$ \cite{Petersan1998a}.  Taking advantage of
the high impedance of the resonator, our DQD-resonator system achieves a
coupling strength $g_\text{c} /2\pi \approx\SI{70}{\MHz}$ \cite{Wang2021a} at resonance
$\Delta E/h=\nu_\text{r}$ for $2t_\text{c}/h =\SI{6.2}{\GHz}$, where
$h$ is the Planck constant. The coupling strength is larger than both qubit
decoherence rate $\gamma /2\pi \approx \SI{50}{\MHz}$ and
photon loss rate $\kappa/2\pi \approx\SI{65.2}{\MHz}$, indicating that
the cQED system operates in the strong coupling limit
\cite{Stockklauser2017a, Mi2017b, Bruhat2018a}, which enables a photonic
probe with better resolution than conventional \SI{50}{\ohm} resonators.

\begin{figure}[thp]
\includegraphics[width=\columnwidth]{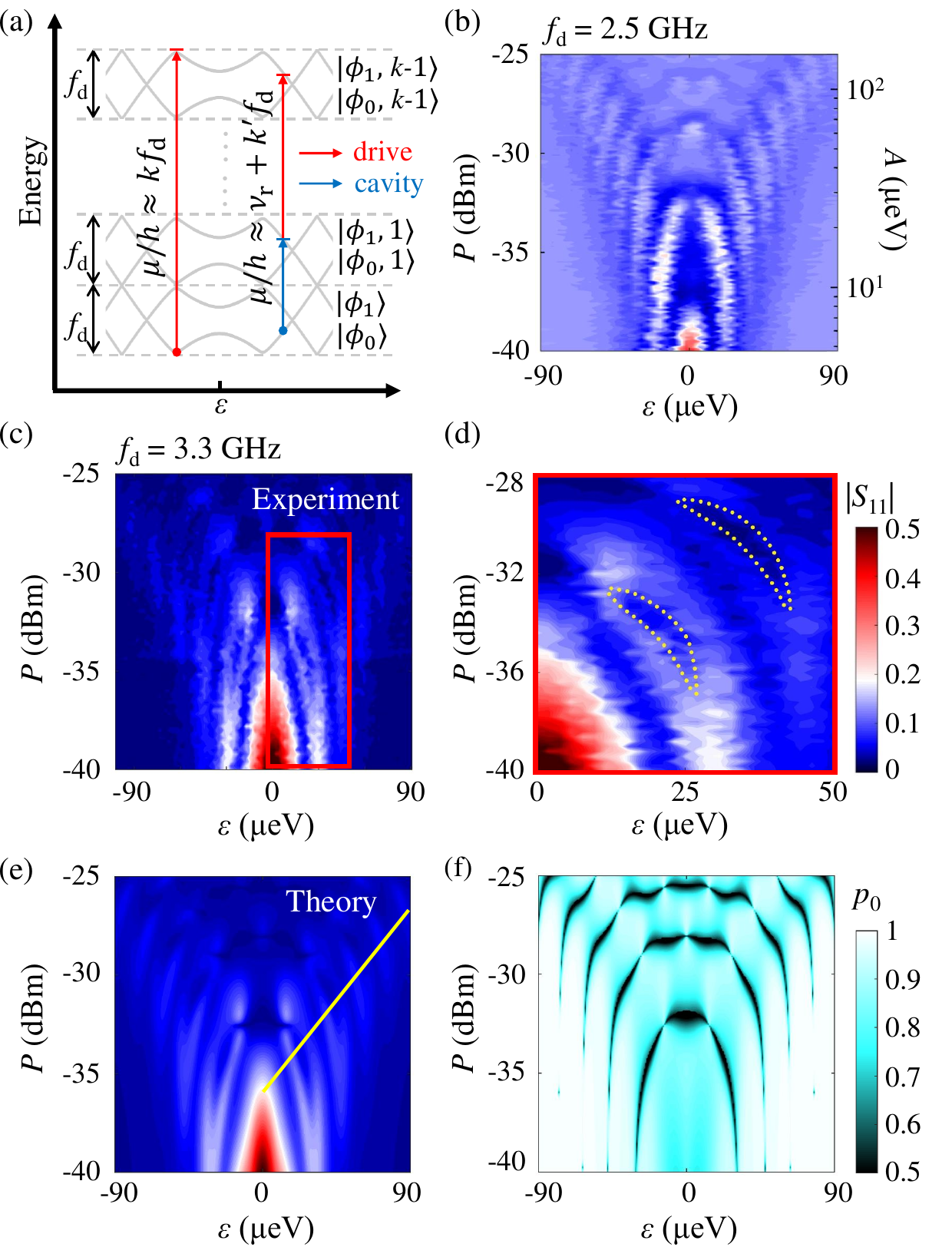}
\caption{
(a) Schematic energy level diagram explained by Floquet theory.
The quasienergy spectrum of the Floquet states has a Brillouin zone structure, 
which is defined by multiples of the energy quantum of the driving field, $hf_\text{d}$. 
The state transfer between the two Floquet states, 
$|\phi_0 \rangle$ and $|\phi_1,k\rangle$, 
can be realized with the population resonance condition (red arrow) 
or cavity-assisted LZSM resonance condition (blue arrow plus red arrow). 
(b)--(c) LZSM interference for different drive frequencies $f_\text{d}$. Resonator reflectance $|S_{11}|$ measured as a function of detuning  $\varepsilon$
and drive power $P$ for $f_\text{d}=\SI{2.5}{\GHz}$ (b),
\SI{3.3}{\GHz} (c), and $2t_c/h \approx\SI{6.2}{\GHz}$.
(d) A zoom-in view of panel (c). The crescent-like holes are outlined using
yellow dotted lines.
(e) Simulated result for the parameters used in panel (c), where $P=\SI{-29}{dBm}$
corresponds to $A=\SI{60}{\mu\eV}$.
To consider an inhomogeneous broadening of the LZSM interference pattern
induced by the slow fluctuations of the detuning $\varepsilon$
\cite{Cao2013a,Forster2014a,Mi2018b}, all theoretical plots are
convoluted with a Gaussian of width $\SI{2.5}{\micro\eV}$.
The yellow line marks the parameters used in Fig.~\ref{fig:spectrum}.
(f) Calculated mean population of the Floquet state $|\phi_0\rangle$.
}
\label{fig:crescents}
\end{figure}

We apply a continuous microwave modulation to the left
barrier gate L, such that the interdot detuning acquires a time dependence
$\varepsilon \to \varepsilon + A\cos(2\pi f_\text{d} t)$, where $A$ is the microwave amplitude, $f_\text{d}
= 1/T$ and $T$ is the period of the driving.  The system can thus be described by Floquet theory
\cite{Shirley1965a}, see Appendix~\ref{app:theory}, which states that the
Schr\"odinger equation has solutions of the form $e^{-2\pi i\mu_\alpha t/h}
|\phi_\alpha(t)\rangle$, $\alpha=0,1$, where $|\phi_\alpha(t)\rangle$ are
the Floquet states satisfying $|\phi_\alpha(t+T)\rangle = |\phi_\alpha(t)\rangle$.  The phase factor is
governed by the quasienergy $\mu_\alpha$, which is distinct from the mean
energy defined as the time-average of the energy expectation value, $E_\alpha
\equiv \frac{1}{T}\int_0^T \, \mathrm{d} t\langle\phi_\alpha(t)|H(t)|\phi_\alpha(t)\rangle$.
The Floquet states can be
Fourier decomposed into components separated by the energy quantum of the
driving field, $hf_\text{d}$, as sketched in Fig.~\ref{fig:crescents}(a).
The quasienergy splitting $\mu \equiv \mu_1-\mu_0$ determines the relative
phase $2\pi\mu T/h$ between two Floquet states acquired during one driving period. 
When this phase is close to a multiple
of $2\pi$, i.e., when $\mu/h \approx kf_\text{d}$ with an integer $k$, the two Floquet
states will interfere constructively, so that both
become significantly populated \cite{Shevchenko2010a}.  
Consequently, LZSM patterns appear in conductance or electron occupation when such a driven
DQD is probed via transport or charge sensing \cite{Stehlik2012a,
Forster2014a, Bogan2018a}.

The driven dynamics of our DQD is probed by the connected microwave
resonator \cite{Blais2004a}.  Specifically, we send in a microwave tone
via the resonator and measure its reflection.
We choose the probe microwave to be at
resonance with the resonator, so that the coupling between the DQD and the
resonator causes a dispersive frequency shift $\nu_\text{r} \to
\nu_\text{r} + (g_\text{c}^2/2\pi)\chi^{(0)}(\nu_\text{r})$ \cite{Kohler2018a},
which modifies the reflection.  The central quantity here is the
phase-averaged susceptibility $\chi^{(0)}$ of the DQD dipole operator
\cite{Kohler2018a},
\begin{equation}
\chi^{(0)}(\nu) = (p_0-p_1) \sum_{k}
\frac{|Z_{10,k}|^2}{2\pi \nu-2\pi\mu/h+2\pi kf_\text{d}+i\gamma/2} ,
\label{chi}
\end{equation}
where $Z_{10,k}$ is the $k$th Fourier component of the transition matrix
element $\langle\phi_0(t)|\sigma_z|\phi_1(t)\rangle$, and $p_\alpha$ is the
occupation probability of the Floquet state $|\phi_\alpha \rangle$ in the
stationary limit.  The $p_\alpha$ are obtained as the well-defined
stationary solution of a Bloch-Redfield master equation which allows a
realistic description of the weakly dissipative qubit.  Details and full
expressions can be found in Appendix~\ref{app:theory}.

According to Eq.~\eqref{chi}, $\chi^{(0)}(\nu)$ and hence the
cavity reflection assume an appreciable size when two conditions are
fulfilled.  
First, with the probe frequency set at the resonator frequency
$\nu=\nu_\text{r}$, the denominator yields the resonance condition $\mu/h =
\nu_\text{r} + k'f_\text{d}$ for some integer $k'$. 
It resembles the resonance condition for the population, but contains an additional cavity photon.  
The modified fringes observed in the reflection can thus be interpreted as cavity-assisted LZSM interference with one state
dressed by a cavity photon \cite{Koski2018a, Kohler2018a, Mi2018b,
Shevchenko2018a}.  
Second, one Floquet state must be predominantly
populated such that $|p_0-p_1|\approx 1$.
Figure~\ref{fig:crescents}(b) shows such Floquet spectroscopy measured by
the reflectance of the resonator $|S_{11}|$ as a function of interdot
detuning $\varepsilon $ and drive power $P\propto A^2$ for $f_\text{d} =
\SI{2.5}{\GHz}$ and $2t_\text{c}/h \approx\SI{6.2}{\GHz}$.  As expected,
the LZSM pattern is similar to the one reported in Ref.~\cite{Koski2018a}.

An interesting and qualitatively new situation arises when the resonance
conditions for both the transition between Floquet states and the
cavity-assisted transition coincide, i.e., when the system reaches the
double resonance
\begin{equation}
kf_\text{d} = \mu/h = \nu_\text{r} + k'f_\text{d} \,.
\end{equation}
This implies that the drive frequency obeys $f_\text{d} = \nu_\text{r}/2,
\nu_\text{r}/3,\ldots$ ($f_\text{d} = \nu_\text{r}$ is discarded because it
would directly affect the cavity and 
pollute the measurement).

Figure \ref{fig:crescents}(c) shows the LZSM pattern for the ``one-half
resonance'' $f_\text{d}=\SI{3.3}{\GHz} \approx \nu_\text{r}/2$.  Compared
to the cavity-assisted LZSM pattern in Fig.~2(b), every fringe splits up
and acquires a crescent-shaped hole, where the reflection is strongly
suppressed.  The emergence of these ``crescents'' can be attributed to the strong driving effect on the Floquet state occupation that modifies the cavity-assisted LZSM interference patterns.

The impact of Floquet state occupation dynamics can be seen in the expression of the susceptibility \eqref{chi}. While the
denominators give the cavity-assisted LZSM resonance conditions, the
occupation probability $p_\alpha$ also plays an important role.
When the resonance condition for the transition between the Floquet states, $\mu/h
= kf_\text{d}$, is fulfilled, the constructive interference
between the Floquet states leads to a significant population of both states,
such that $p_1\approx p_0=1/2$ and $p_0 - p_1 \approx 0$. Consequently, $\chi^{(0)}$ is strongly suppressed, and the
cavity signal diminishes.  A numerical simulation based on Eq.~\eqref{chi} is shown in
Fig.~\ref{fig:crescents}(e), which is in good agreement with the experiment.
The expected population of the Floquet
state with smaller quasienergy in Fig.~\ref{fig:crescents}(f),
confirms that the crescents emerge when the population of this state is
significantly reduced.

\begin{figure}[tp]
\centerline{\includegraphics[width=1.2\columnwidth]{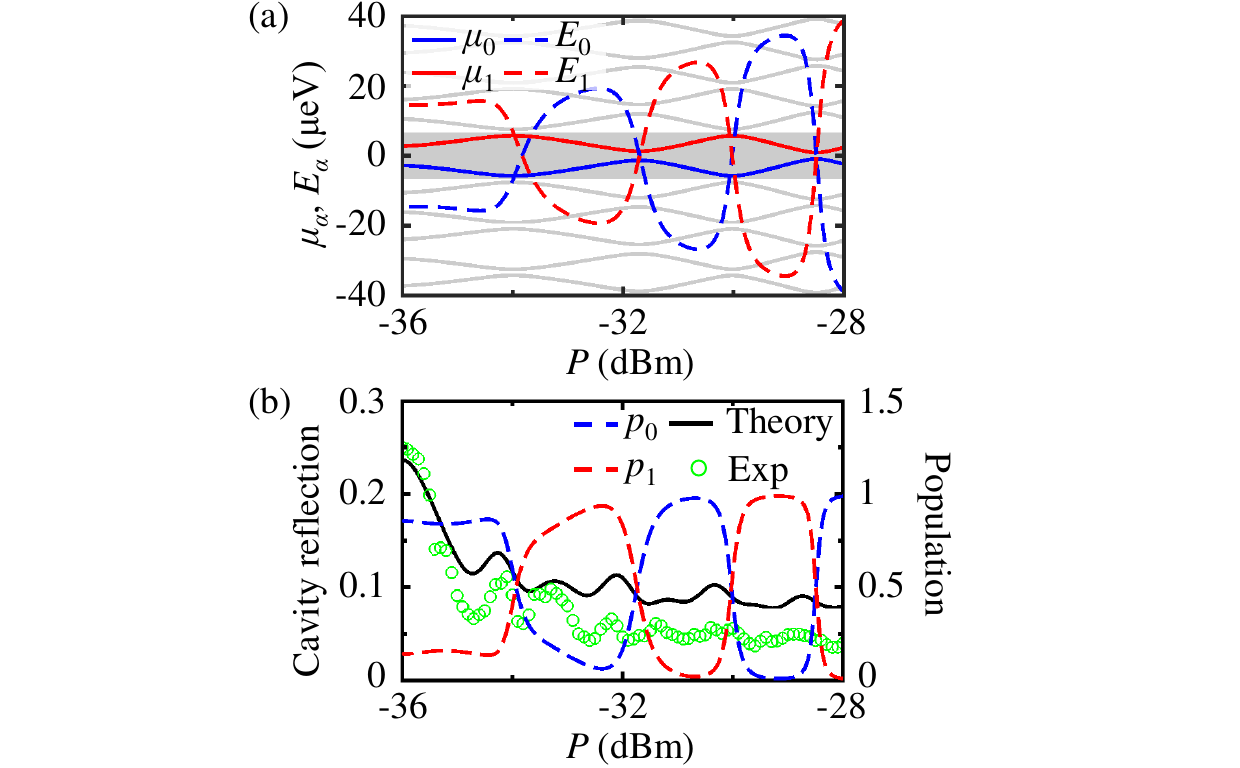}}
\caption{(a) Floquet spectrum (solid) in the first Brillouin
zone (shaded area) and corresponding mean energies (dashed).
Gray lines mark the sidebands of the quasienergies.
The microwave power and the
detuning are both varied along the line in Fig.~\ref{fig:crescents}(e).
(b) Corresponding populations in the stationary state (dashed) together
with the measured (circles) and the simulated (solid) cavity response.}
\label{fig:spectrum}
\end{figure}

\begin{figure*}[tp]
\includegraphics[width=17.8cm]{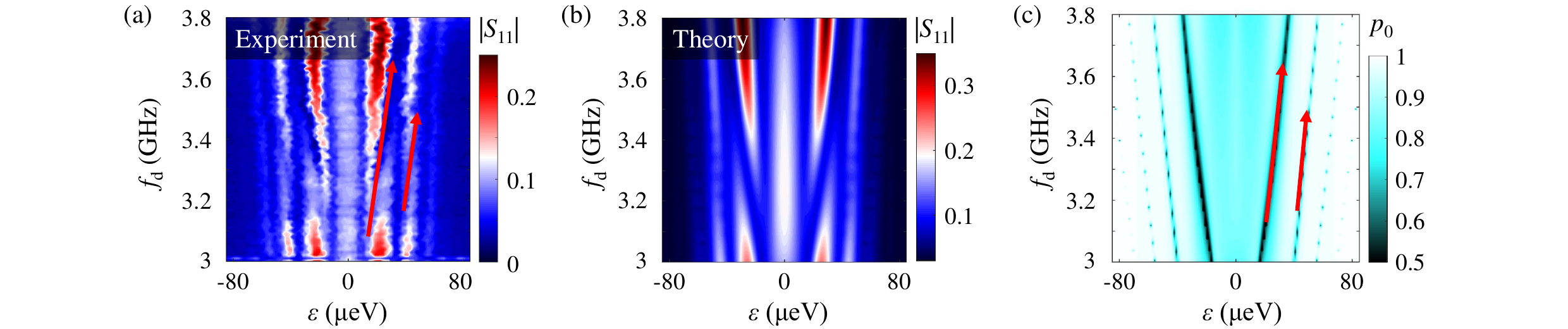}
\caption{(a) Resonator reflectance $|S\textsubscript{11}|$ as a function of
detuning $\varepsilon$ and drive frequency $f_\text{d}$ for power
$P = \SI{-36}{dBm}$ and $2t/h=\SI{6.2}{\GHz}$. The evolution of the crescents is indicated by red arrows.
(b) Numerical simulation for the parameters used in panel (a).  (c) Corresponding
mean population of the Floquet state $|\phi_0\rangle$.}
\label{fig:freq}
\end{figure*}

The analysis above provides an intuitive and qualitative picture
for the emergence of the crescents in the cavity-assisted LZSM patterns, and points to the importance of the Floquet state spectrum and population. 
We thus examine them more closely, and calculate the mean energies and populations of the Floquet states.
The solid lines in Fig.~\ref{fig:spectrum}(a) show the quasienergies
$\mu_{0,1}$ and their replica for parameters marked by the line in Fig.~\ref{fig:crescents}(e).
Most significant are the avoided quasienergy
crossings located at the center and border of the Brillouin zone.
Since the crossings are avoided, the resonance
condition for the population, $\mu = khf_\text{d}$, will only be fulfilled
approximately.  At the avoided quasienergy crossings,
the mean energies generically cross exactly, see Appendix~\ref{app:theory}.

The key to the cavity response, however, is not whether $\mu = khf_\text{d}$ can be satisfied exactly, but how the populations of the Floquet states behave.  
As shown in Fig.~\ref{fig:spectrum}(b),  the Floquet state with lower mean energy is populated predominantly.
In this case the prefactor of the susceptibility \eqref{chi} is of the order unity.  
However, at the avoided quasienergy crossings, the populations in both Floquet states are equal, $p_{0} = p_1 = 1/2$.  
Consequently, the DQD susceptibility vanishes and the simulated cavity reflection assumes its minimum. 
In comparison, the corresponding experimental result is plotted with circles in Fig.~\ref{fig:spectrum}(b).
It shows good agreement for the positions of the minima.
The magnitude of the reflection, including the visibility of the fringes, is predicted fairly well, too.
For drive power beyond the second crossing, the agreement becomes more qualitative
because the heating effect caused by the large power deteriorates the system coherence.
Nevertheless, the agreement between experiment and simulation clearly shows that dispersive readout by a cavity can be employed to probe changes of the stationary Floquet state population and quasienergy crossings.

With cavity frequency $\nu_\text{r}$ fixed, the double resonance conditions can be satisfied when sweeping both the drive frequency $f_\text{d}$ and the qubit detuning $\varepsilon$.
As discussed above, we expect a peak in the cavity reflection for $\mu/h =
\nu_\text{r} + k'f_\text{d}$, while a valley emerges for $\mu/h = kf_\text{d}$.
With $\mu/h = \nu_\text{r} + k'f_\text{d} = kf_\text{d}$ ($k\neq k'$), we expect that as a function of $f_\text{d}$,
the peaks and valleys in the cavity reflection in the spectroscopy have different inclinations and cross each other.

Figure \ref{fig:freq}(a) depicts this feature of the measured reflection as a function of
$\varepsilon$ and $f_\text{d}$ in the vicinity of the one-half resonance
$\nu_\text{r}/2$ at a fixed driving power.  As expected, upon increasing
$f_\text{d}$, the gaps (signifying suppression of reflection) in the cavity
fringes evolve outwards and intersect the interference stripes when
$f_\text{d} = \nu_\text{r}/2$.  A selection of the corresponding LZSM
patterns is shown in Appendix~\ref{app:data}.  The simulated measurement in
Fig.~\ref{fig:freq}(b) is in good agreement with the experiment.  The
reduced population of the Floquet state $|\phi_0\rangle$ can be identified
as the origin of the collapsing cavity signal.  We also computed the
expected population of the Floquet state, as shown in
Fig.~\ref{fig:freq}(c).  Indeed, along the gaps in the interference strips,
$p_0$ is reduced to a value considerably below unity.  In short,
Fig.~\ref{fig:freq} not only reconfirms the interplay between the two
resonances, but also demonstrates how we can influence and electrically control
the dynamics of Floquet states.

With our high-impedance superconducting resonator we were able to measure a
crescent-like pattern near the one-half resonance
$f_\text{d}=\nu_\text{r}/2$.  
Theoretically, one would expect to find a similar structure for $f_\text{d} = \nu_\text{r}/3$, $\nu_\text{r}/4,\ldots$  
However, the LZSM fringes in the populations will be narrow with decreasing drive frequency \cite{Shevchenko2010a} 
such that the narrow crescents are blurred by the typical inhomogeneous broadening in GaAs \cite{Petersson2010a, Cao2013a}.
Nevertheless, we could still observe the characteristic feature of the
one-third resonance, which appears as claw-shaped patterns at the end of
the interference fringes as shown in Fig.~\ref{fig:third} in Appendix~\ref{app:data}.

In conclusion, we have demonstrated a novel interference pattern in the
Floquet spectroscopy of a driven semiconductor DQD probed by a
high-impedance resonator.  Specifically, when the driving microwave tone is
at half of the resonator frequency, crescent-shaped holes/depressions
appear in the LZSM interference peaks of the cavity reflection.
The physical origin can be explained by the reduced
cavity response due to the population redistribution of Floquet states.
A systematic
investigation revealed that the resonance condition for the population is
approximately fulfilled at avoided quasienergy crossings at which the
stationary state of the DQD turns from an almost pure state into a mixture.
Therefore the emergence of the crescents can be considered as an
experimental signature of the nature of the stationary state of a strongly
driven two-level system.  The LZSM patterns emerging from double resonance
are not particular to semiconductor-resonator hybrid systems, but should be
generic for all cQED systems.  They may be used to gain further insight
into Floquet state dynamics in the presence of strong driving.

\begin{acknowledgments}
This work was supported by the National Key Research and Development
Program of China (Grant No.\ 2016YFA0301700), the National Natural Science
Foundation of China (Grants No.\ 61922074, 11674300, 61674132, 11625419 and
11804327), the Strategic Priority Research Program of the CAS (Grant No.\
XDB24030601), the Anhui initiative in Quantum Information Technologies
(Grant No.\ AHY080000), the Spanish Ministry of Science, Innovation, and
Universities (Grant No.\ MAT2017-86717-P) and by the CSIC Research Platform
on Quantum Technologies PTI-001.  H.-W. J. and X. H. acknowledge financial
supported by U.S. ARO through Grants No.\ W911NF1410346 and No.\
W911NF1710257, respectively.  This work was partially carried out at the
University of Science and Technology of China Center for Micro and
Nanoscale Research and Fabrication.
\end{acknowledgments}

\appendix

\section{Additional data}
\label{app:data}

\begin{figure}[b]
\includegraphics[width=\columnwidth]{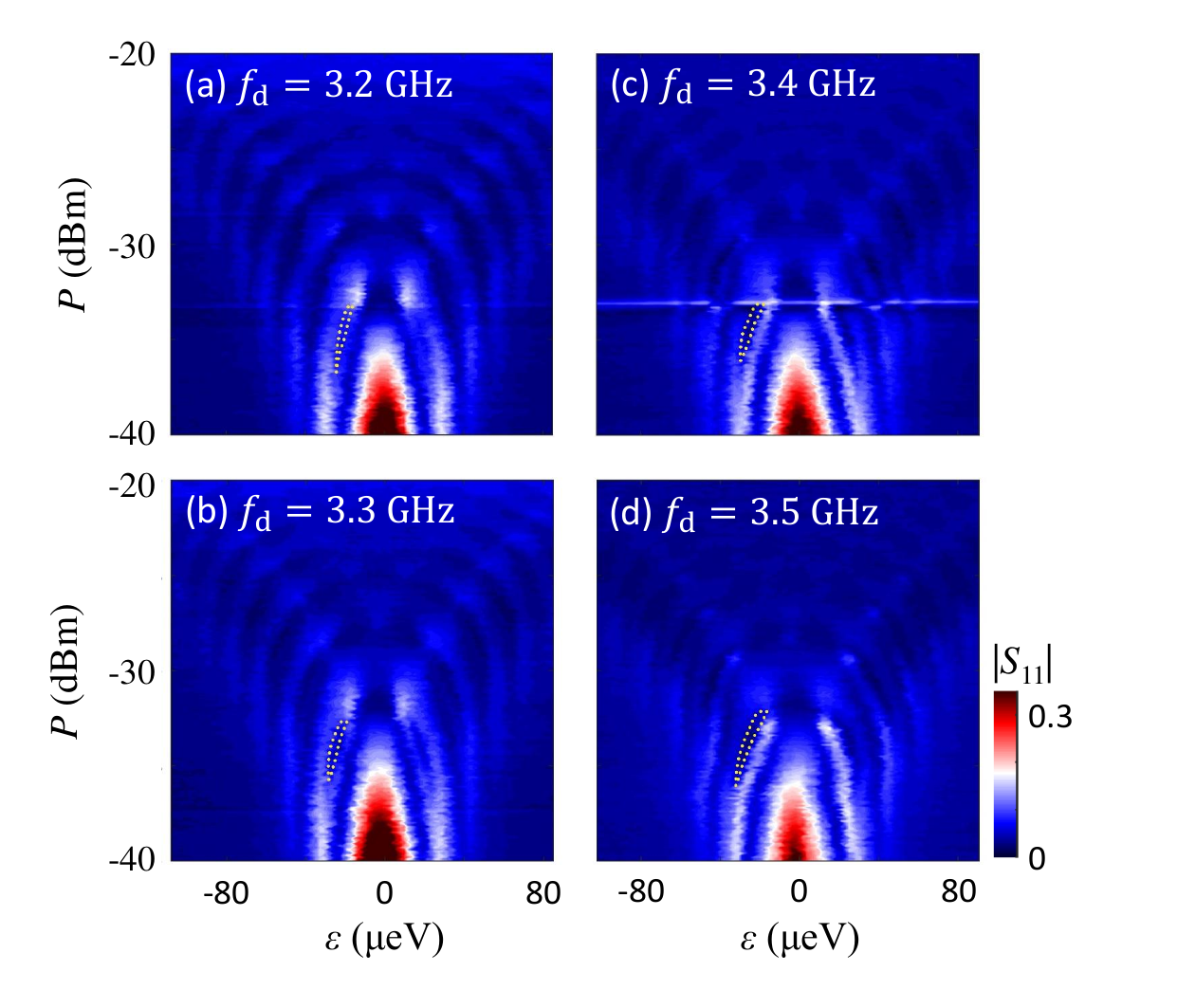}
\caption{LZSM interference patterns for various drive frequencies
$f_\text{d}$ near the half-resonance $\nu_\text{r}/2$ ranging from
\SI{3.2}{\GHz} to \SI{3.5}{\GHz}.  The crescents are outlined by yellow
doted lines.}
\label{fig:DriveTuning}
\end{figure}

\subsection{LZSM patterns for different drive frequencies}

Here we illustrate how the crossing of the bright and the dark stripe [see
Fig.~4 of the main text] manifests itself in the LZSM pattern.
To this end, we slightly vary $f_\text{d}$ around the half-resonance
$\nu_\text{r}/2$.  The resulting measured patterns are shown in
Fig.~\ref{fig:DriveTuning}.  With increasing drive frequency $f_\text{d}$,
the crescent-shaped holes in the fringes move outwards, which correspond
to a shift along the red arrows in Fig.~4(a).

\subsection{Tuning the resonator frequency}

\begin{figure}[b]
\includegraphics[width=\columnwidth]{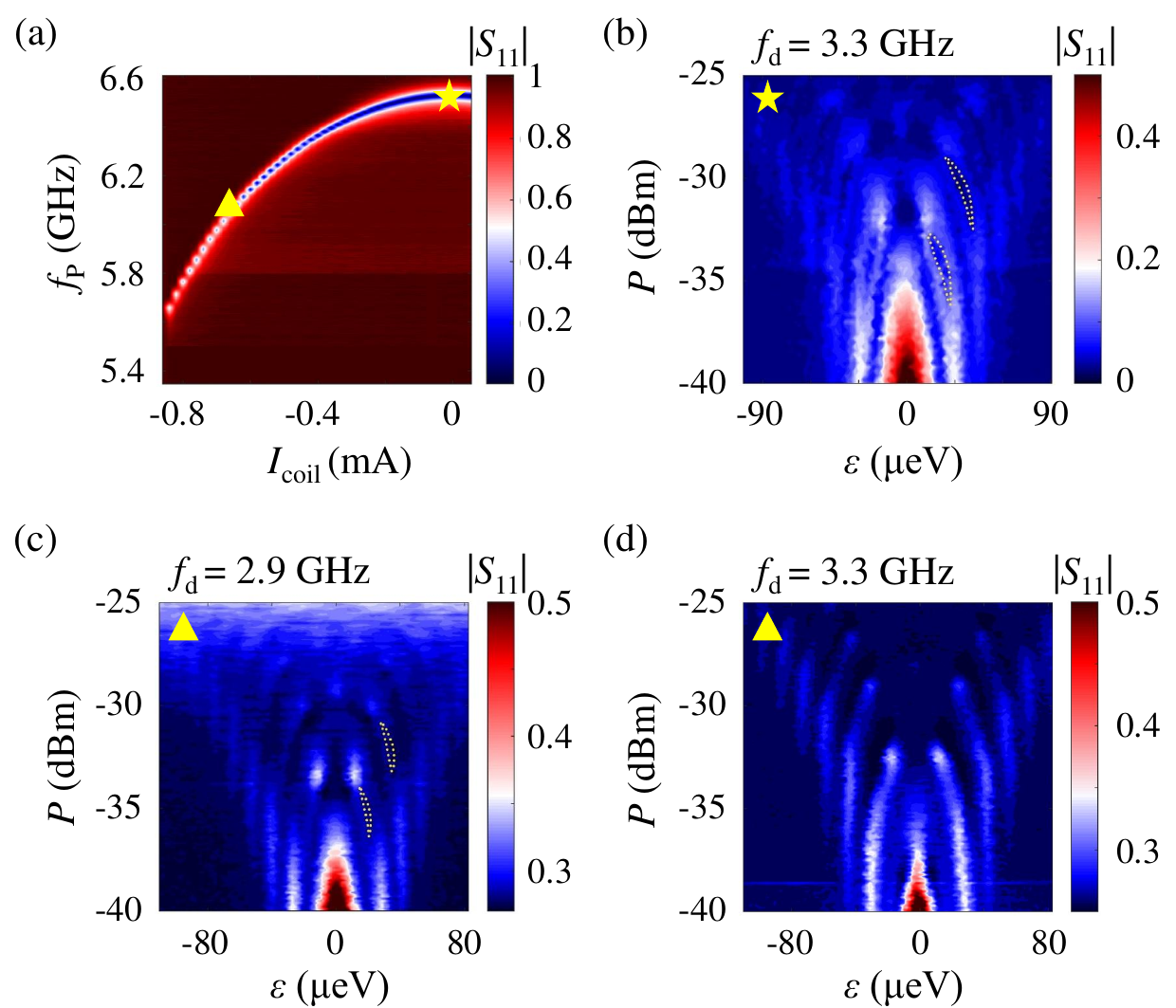}
\caption{(a) Reflectance spectrum $|S_{11}|$ as a function of
probe frequency $\nu_\text{p}$ and current $I_\text{coil}$ through the coil.
Resonance frequency $\nu_\text{r}$ is tuned from \SI{6.51}{\GHz} to \SI{6.07}{\GHz}
indicated by a star and a triangle, respectively. (b--d) Interference
patterns for different $\nu_\text{r}$ in (a) and different $f_\text{d}$.}
\label{fig:coil}
\end{figure}

Taking advantage of the tunability of the SQUID array resonator, we are
able to vary also the resonance frequency $\nu_\text{d}$ by the current
$I_\text{coil}$ of a cm-sized coil mounted on the sample holder from
\SI{6.5} {\GHz} to below \SI{5.2}{\GHz} as shown in Fig.~\ref{fig:coil}(a).
Tuning the resonance frequency $\nu_r$ from \SI{6.51}{\GHz} to
\SI{6.07}{\GHz} while keeping the drive frequency at
$f_\text{d}=\SI{3.3}{\GHz}$, the crescents in Fig.~\ref{fig:coil}(d) vanish
once the condition for double resonance, $\nu_\text{r}=nf_\text{d}$ with $n\geq 2$, is sufficiently
violated.  In contrast, when we decrease $f_\text{d}$ to $\SI{2.9}{\GHz}$
which is close to $\nu_\text{r}/2$ (for $f_\text{d}$ a value below one-half resonance 
\SI{3}{\GHz} may result from the shift of the resonance frequency due to the
fluctuation of the magnetic environment), crescents arise again in
Fig.~\ref{fig:coil}(c).  Thus, we have tested our conjecture for the
interplay of cavity reflection and Floquet state population by
independently varying all relevant frequencies.

\subsection{Crescents of higher order}

In the experiment, we have observed crescents only for the one-half
resonance $f_\text{d}\approx \nu_\text{r}/2$. 
For resonances of higher order, i.e., for $f_\text{d}\approx \nu_\text{r}/k$ with $k>2$, similar patterns are expected to be observed.
However, the LZSM fringes in the populations will diminish with decreasing drive frequency \cite{Shevchenko2010a} such that their
size eventually drops below the experimental resolution.

\begin{figure}[b]
\includegraphics[width=\columnwidth]{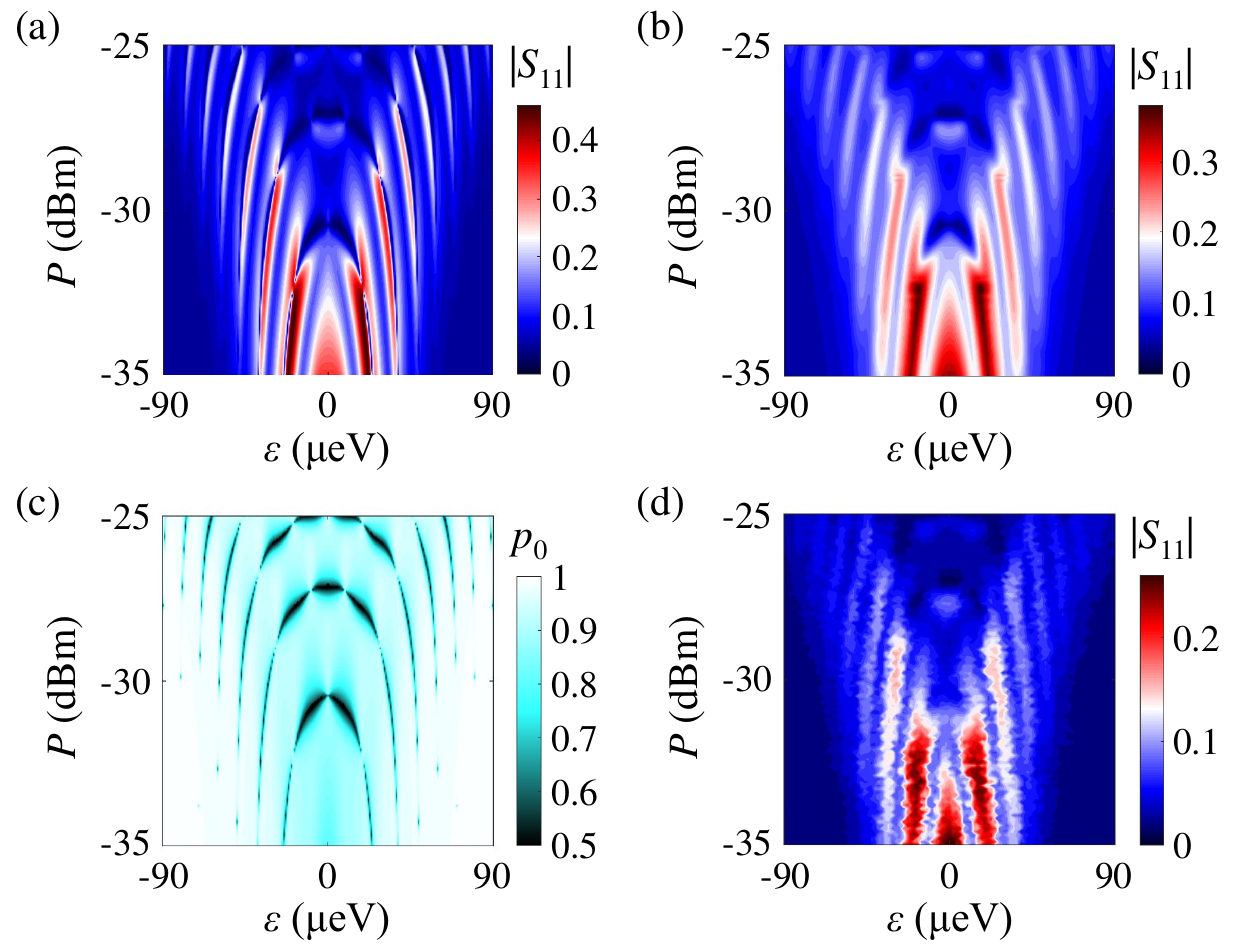}
\caption{
(a) Theoretical prediction for the cavity response for the one-third
resonance with driving frequency $f_\text{d} = \SI{2.27}{\GHz} \approx
\nu_\text{r}/3$ at which the drive power $\SI{-29}{dBm}$ corresponds to
$A=\SI{45}{\micro\eV}$.
(b) The same data, but after considering for the detuning $\varepsilon$ an
inhomogeneous broadening of $\SI{2.5}{\micro\eV}$.
(c) Corresponding mean population of the Floquet state $|\phi_0\rangle$.
(d) Experimental data for the resonator reflectance $|S_{11}|$ with $f_\text{d} = \SI{2.27}{\GHz}$.
}
\label{fig:third}
\end{figure}

We show in Fig.~\ref{fig:third}(a) the theoretical
prediction for the cavity reflection near the one-third resonance, which
exhibits very narrow crescents.  
However, to predict the result of a realistic measurement, we have to take into account that the level detuning
$\varepsilon$ is slowly fluctuating.  
This causes an inhomogeneous broadening of the order of some $\SI{}{\micro\eV}$ \cite{Petersson2010a, Forster2014a}
which can be captured by convolving the theory data with a Gaussian of
corresponding width.  
Figure~\ref{fig:third}(b) depicts the resulting
prediction for a measurement, where we have assumed a
broadening of $\SI{2.5}{\micro\eV}$.  
This blurs the LZSM pattern so much that the narrow crescents practically disappear. 
Nevertheless, we could still find the characteristic feature of the one-third resonance, which appears as claw-shaped patterns at the end of the interference fringes.
Figure~\ref{fig:third}(d) demonstrates these claw-shaped patterns measured in the experiment.

\section{Theoretical background and methods}
\label{app:theory}

In the following, we briefly review the theoretical tools and concepts used
in the main text.

\subsection{Floquet theory}

The Schr\"odinger equation of a $T$-periodically driven quantum system
possesses a complete set of solutions of the form \cite{Shirley1965a}
\begin{equation}
|\psi_\alpha(t)\rangle = e^{-i\mu_\alpha t/\hbar} |\phi_\alpha(t)\rangle ,
\end{equation}
where the Floquet states $|\phi_\alpha(t)\rangle$ obey the periodicity of
the driving, while the quasienergies $\mu_\alpha$ determine the phase
acquired by $|\psi_\alpha(t)\rangle$ during one driving period.  Both
the quasienergies and the Floquet states can be obtained by solving the
Floquet eigenvalue equation
\begin{equation}
\mathcal{H}(t)|\phi(t)\rangle \equiv
\Big( H(t) -i\hbar \frac{d}{dt}\Big)|\phi(t)\rangle
= \mu|\phi(t)\rangle .
\label{app:floquet}
\end{equation}
The quasienergy spectrum has a Brillouin zone structure, which means that
they are defined only up to multiples of the energy quantum of the driving
field, $\hbar\Omega$, where $\Omega = 2\pi/T$ is the angular frequency of
the driving.  As the quasienergies follow from an eigenvalue equation, they
exhibit level repulsion, i.e., in the absence of symmetries they
generically form avoided crossings.  At such avoided crossings the
participating states interchange their physical properties, in particular
their mean energies
\begin{equation}
E_\alpha = \frac{1}{T} \int_0^T dt \,
\langle\psi_\alpha(t)|H(t)|\psi_\alpha(t)\rangle .
\label{app:E}
\end{equation}
This implies that at avoided quasienergy crossings, the corresponding mean
energies cross exactly.

\subsection{Floquet spectrum of the two-level system}

In the main text, we describe the DQD by the two-level Hamiltonian
\begin{equation}
H(t) = \frac{\Delta}{2}\sigma_x
+ \frac{\varepsilon+A\cos(\Omega t)}{2}\sigma_z \,,
\label{app:Ht}
\end{equation}
where $\Delta$ is the interdot tunnel coupling, $A$ is the drive amplitude and $\sigma_{x,y,z}$ are Pauli matrices.  Let us consider the
transformation of the corresponding Floquet Hamiltonian $\mathcal{H} =
H(t)-i\hbar\,d/dt$ under the anti-unitary transformation $\theta = i\sigma_y K$,
where $K$ causes complex conjugation, i.e., $KzK^{-1} = z^*$.  One readily sees
that $\theta\mathcal{H}(t)\theta^{-1} = -\mathcal{H}(t)$ (notice that for
the present reasoning, we do not invert the time argument of the
Hamiltonian), which implies that any solution of the Floquet equation
\eqref{app:floquet} with eigenvalue $\mu$ has a time-reversed partner with
eigenvalue $-\mu$.  It is therefore convenient to choose the Brillouin zone
symmetrically around zero, $-\hbar\Omega/2 < \mu_\alpha \leq
\hbar\Omega/2$, which for the two-level system ensures the relation $\mu_0
+ \mu_1 = 0$.

It is now tempting to conclude that (exact) quasienergy crossings can occur
only when $\mu_0 = \mu_1 = 0$.  This however, would ignore crossings with
equivalent states from other Brillouin zones with quasienergies shifted by
multiples of $\hbar\Omega$.  Taking them into account yields for the
crossings the weaker condition $\mu_0 + k\hbar\Omega = \mu_1$, where
$k$ may be any integer, such that $\mu_0 = k\hbar\Omega/2$.  This means
that for the Hamiltonian \eqref{app:Ht}, quasienergies may cross only at
the border or in the middle of the Brillouin zone.  In the absence of
symmetries, the crossings are generally avoided, but nevertheless are found
	under the same condition, see Fig.~\ref{fig:spectrum}(a).

\subsection{Stationary state of a driven system}

In quantum mechanics, dissipation can be modeled by coupling the system
to a heat bath. 
For weak dissipation, one may
eliminate the bath by second order perturbation theory to obtain a
Markovian master equation for the reduced density operator.  For a driven
system, Floquet states form a suitable basis for the numerical treatment of
such master equations, because then the driving is already included by the
choice of the basis \cite{Kohler1997a}.  Finally, one obtains for the
populations $p_\alpha$ of the Floquet states a rate equation of the form
\begin{equation}
\dot p_\alpha = \sum_\beta ( w_{\alpha\leftarrow\beta} p_\beta
- w_{\beta\leftarrow\alpha} p_\alpha) .
\label{app:ME}
\end{equation}
The transition rates $w_{\alpha\leftarrow\beta}$ are given by
golden-rule like expressions for the transition matrix elements of the
operator that couple the system to the bath.  In the numerical
calculations of the main text, this coupling is established via the qubit
operator $\sigma_x$.  This choice leads to the observed generic
LZSM pattern \cite{Blattmann2015a} and fits the observation rather well.

Generally, this master equation possesses a unique stationary solution with
$\dot p_\alpha=0$ for all $\alpha$, despite that it does not fulfill a
detailed balance condition.  The latter makes it virtually impossible to
find a generic formal solution for the stationary state.  Nevertheless, in
a few cases this is possible.  Here we sketch two typical limits and apply
the ideas to the driven two-level system described by the Hamiltonian
\eqref{app:Ht}.

\begin{figure}
\includegraphics{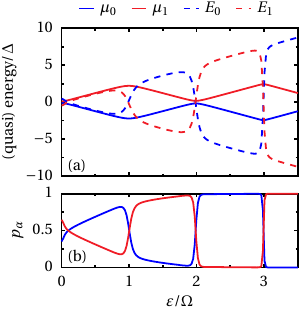}
\caption{(a) Quasienergies (solid lines) and mean energies (dashed) for the
driven two-level system as function of the detuning for driving frequency
$\Omega=5\Delta$ and amplitude $A=2\Omega$ at zero temperature in the weak
coupling limit.  The bath couples to the system operator $\sigma_x$.
(b) Populations of the corresponding Floquet states in the stationary limit.}
\label{fig:ME}
\end{figure}

\begin{figure}
\includegraphics{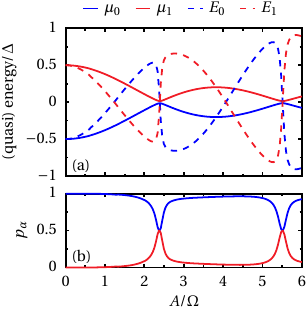}
\caption{The same as Fig.~\ref{fig:ME} for detuning
$\varepsilon=0.03\Delta$ as function of the driving amplitude for a bath
coupled via $\sigma_z$, which far from the quasienergy crossings leads to a
Floquet-Gibbs state.}
\label{fig:FG}
\end{figure}

For small or intermediate frequencies, a natural guess is that the system
eventually resides in the Floquet state with the smallest mean energy,
possibly with a small admixture of other Floquet states.  This was indeed
found for systems with mixed chaotic/regular phase space with some exception
close to avoided crossings \cite{Kohler1998a}.  For the present case of a
two-level system, the spectra and the populations are shown in
Fig.~\ref{fig:ME}.  This confirms that far from the crossings, the state
with smaller mean energy is predominantly occupied.  At the center of the
crossing, however, both states have equal mean energy and, consequently, we
observe a mixture of both Floquet states, where each one is occupied with
probability $p_0 \approx p_1 \approx 1/2$.

For high-frequency driving, the system may be mapped approximately to a
time-independent Hamiltonian \cite{Holthaus1992b, Grossmann1992a}.  If in
addition, the system-bath coupling commutes with the driving, one often
finds a Floquet-Gibbs state, i.e., a statistical mixture that resembles a
Gibbs states but with the energies replaced by quasienergies
\cite{Hartmann2017a, Engelhardt2019a}, provided that the Brillouin zone is
chosen such that it covers the spectrum of the undriven system.  For the
two-level system, this was found for zero detuning \cite{Engelhardt2019a}.
For small detuning this is still valid, as can be appreciated in
Fig.~\ref{fig:FG}.  In panel (b), we indeed see that, in contrast to
Fig.~\ref{fig:ME}(b), the populations are governed by the quasienergies.

These situations emphasize the role of avoided quasienergy crossings for
the long-time solution of a driven dissipative quantum system.  The
characteristic feature is that far from the crossings, one Floquet state
dominates, while at the crossings, both states are equally populated.  In
our experiment, these equal populations are responsible for the
emergence of crescents in the LZSM patterns.

\subsection{AC-driving and dispersive readout}

A Floquet theory for the response of a microwave cavity coupled to a
periodically driven double quantum dot (DQD) has been derived in
Ref.~\cite{Kohler2018a}.  Its cornerstones are summarized in the following.
The central idea of dispersive readout is that a resonantly driven cavity
acts via a coupling $H_\text{coupl} = g_\text{c} Z(a+a^\dagger)$ with some
system operator $Z$ and the creation and annihilation operators $a^\dagger$ and
$a$ of a cavity photon on the DQD (or any other quantum system).
In turn the DQD acts back to the cavity, which leads to a frequency shift
of the cavity that provides information on the DQD state.  Within
linear-response theory, this backaction is captured by the susceptibility
$\chi(t,t') = -i\langle[Z(t),Z(t')]\rangle\theta(t-t')$ with $\theta$ the
Heaviside step function and $Z=\sigma_z$ the operator that couples
the DQD to the cavity.  If the DQD experiences an additional external
driving, the susceptibility depends explicitly on both times.  In the case
of periodic driving, after a transient stage, $\chi(t,t') =
\chi(t+T,t'+T)$, which implies that $\chi(t,t-\tau)$ with $\tau=t-t'$ is
$T$-periodic in $t$.

The susceptibility $\chi$ allows one to calculate via input-output theory
\cite{Collett1984a, Blais2004a} the reflection of our one-sided cavity to
read
\begin{equation}
S_{11} =
\frac{a_\text{out}}{a_\text{in}} = 1 + \frac{i \kappa_{\text i}}
{\omega_c - \omega + g_\text{c}^2 \chi^{(0)}(\omega) - i\kappa / 2} \,,
\label{app:io}
\end{equation}
where $\kappa_{\text{i,e}}$ are the internal and external loss rates of
the cavity, while the total loss $\kappa =\kappa_\text{i}+\kappa_\text{e}$.

The impact of the DQD is contained in the $k=0$ component of the Fourier
transformed susceptibility
\begin{equation}
\chi^{(k)}(\omega) = \int^{\infty}_0  \left[\frac{1}{T}\int^T_0
e^{i k\Omega t+i \omega t} \chi(t,t - \tau) d t \right] \! d\tau ,
\end{equation}
where for $k=0$ the $t$-integration corresponds to the average over one
driving period which is equivalent to averaging over the phase of the
driving.
When the DQD is described by a density operator $\rho_\infty
= \sum_\alpha p_\alpha |\phi_\alpha(t)\rangle\langle\phi_\alpha(t)|$ with
probabilities $p_\alpha$ computed as described above, one obtains for the
susceptibility after some algebra the expression
\cite{Kohler2018a}
\begin{equation}
\label{app:chi0w}
\chi^{(0)}(\omega) = \sum_{\alpha,\beta,k}
\frac{(p_\alpha-p_\beta)|Z_{\alpha\beta,k}|^2}
{\omega+\mu_\alpha-\mu_\beta-k\Omega+i\gamma_{\alpha\beta}/2}.
\end{equation}
Here, $Z_{\alpha\beta,k}$ denotes the $k$th Fourier component of the
$T$-periodic transition matrix element $Z_{\alpha\beta}(t) =
\langle\phi_\alpha(t)|Z|\phi_\beta(t)\rangle$, while the dephasing rates
$\gamma_{\alpha\beta}$ have been introduced phenomenologically.

\bibliography{literature}

\end{document}